\documentclass[12pt,preprint]{aastex}
\usepackage{natbib}
\usepackage{amsfonts}
\usepackage{color}
\usepackage{ulem}
\usepackage{amsmath}
\usepackage{mathtools}
\usepackage{graphicx}
\usepackage[export]{adjustbox}

\newcommand{\rsun}{$\,R_{S}$}
\newcommand{\degrees}{$^{\circ}$}

\shorttitle{Modeling Solar Coronal Shock Proton Acceleration}
\shortauthors{K. A. Kozarev, N. A. Schwadron}

\begin{document}

\title{A Data-Driven Analytic Model for Proton Acceleration by Large-Scale Solar Coronal Shocks\\}

\author{Kamen A. Kozarev\altaffilmark{1}, Nathan A. Schwadron\altaffilmark{2}}
\altaffiltext{1}{Smithsonian Astrophysical Observatory}
\altaffiltext{2}{Institute for the Study of Earth, Oceans, and Space, University of New Hampshire}

\begin{abstract}We have recently studied the development of an eruptive filament-driven, large-scale off-limb coronal bright front (OCBF) in the low solar corona (Kozarev et al. 2015), using remote observations from Solar Dynamics Observatory's Advanced Imaging Assembly EUV telescopes. In that study, we obtained high-temporal resolution estimates of the OCBF parameters regulating the efficiency of charged particle acceleration within the theoretical framework of diffusive shock acceleration (DSA). These parameters include the time-dependent front size, speed, and strength,  as well as the upstream coronal magnetic field orientations with respect to the front's surface normal direction. Here we present an analytical particle acceleration model, specifically developed to incorporate the coronal shock/compressive front properties described above, derived from remote observations. We verify the model's performance through a grid of idealized case runs using input parameters typical for large-scale coronal shocks, and demonstrate that the results approach the expected DSA steady-state behavior. We then apply the model to the event of May 11, 2011 using the OCBF time-dependent parameters derived in Kozarev et al. (2015). We find that the compressive front likely produced energetic particles as low as 1.3 solar radii in the corona. Comparing the modeled and observed fluences near Earth, we also find that the bulk of the acceleration during this event must have occurred above 1.5 solar radii. With this study we have taken a first step in using direct observations of shocks and compressions in the innermost corona to predict the onsets and intensities of SEP events.
\end{abstract}

\section{Introduction}
\label{s1}
Coronal Mass Ejections (CMEs) are massive expulsions of heated, magnetized gas from the tenuous solar atmosphere, the corona. Caused by the catastrophic release of magnetic energy stored in the twisted coronal loops of active regions, which is triggered by magnetic field reconnection, CMEs frequently gain enough energy to leave the Sun's atmosphere at speeds that may exceed 2000 km/s \citep{Gopalswamy:2009}. Within two solar radii from the solar surface, the local Alfv{\'e}n speeds ($V_A=\frac{B}{\sqrt{\mu_0\rho}}$) may drop below 1000 km/s, so CMEs are often capable of driving shock waves low in the corona \citep{Evans:2008, Zucca:2014}. These have been indirectly detected from timings of radio type II emission spectra \citep{Gopalswamy:2013}, as well as in imaging radio observations \citep{Carley:2013}. In the last ten years, the high cadence imaging capabilities of space instruments, such as the Sun Earth Connection Coronal and Heliospheric Investigation \citep[SECCHII]{Howard:2008} on the STEREO mission \citep{Kaiser:2008}, and the Advanced Imaging Assembly \citep[AIA]{Lemen:2012} on board the Solar Dynamics Observatory  mission \citep[SDO]{Pesnell:2012}, combined with Earth- and space-based radio instruments have significantly increased the amount of information and knowledge about these phenomena \citep{Veronig:2010, Bein:2011, Rouillard:2012, Long:2011}. However, the direct connection between the detailed observations of shock dynamics in early-stage solar eruptions and the particle acceleration has remained largely unexplored. Given the current lack of in situ observations in the low solar corona, this connection can be best revealed through modeling of the acceleration process driven by remote observations of multiple events.

\citet{Kozarev:2015} recently studied the development of a large-scale off-limb coronal bright front (OCBF) in the low corona (1.0--2.0\rsun) of the Sun by using remote observations from AIA, combined with several data-driven models of the magnetic field and the change in coronal density. Similar to previous studies \citep{Kozarev:2011,  Downs:2012}, \citet{Kozarev:2015} determined that the observed feature is a driven magnetohydrodynamic (MHD) wave, which steepens into a shock within the AIA field of view (FOV). They obtained estimates of parameters of the OCBF, which regulate the efficiency of acceleration of charged particles within the theoretical framework of Diffusive Shock Acceleration (DSA). These parameters include the time-dependent shock radius $R_{sh}$, speed $V_{sh}$, and strength $r$, as well as the upstream (in the shock frame) potential coronal magnetic field orientations with respect of the shock surface normal, $\theta_{BN}$. Because of the very high cadence of the AIA telescope, we were able to obtain estimates of these quantities for every 12 seconds of the approximately 8 minutes, which the OCBF spent in the AIA field of view. To obtain the time-dependent $\theta_{BN}$ values at multiple locations on the front, we developed the following method: 1) we used a spherical geometric surface model to fit the global shape of the front at consecutive observation times; 2) we computed the global potential coronal field for the corresponding time using a Potential Field Source Surface (PFSS) model \citep{Schrijver:2003}; 3) at each observation time, we determined the locations where individual coronal magnetic field lines intersected the fitted spherical shock surface, and calculated the upstream $\theta_{BN}$ and magnetic field magnitude, $|B|$. The density compression ratios were obtained by applying a Differential Emission Measure (DEM) model \citep{Aschwanden:2013} to the EUV observations, and calculating the emission measure ratios before and during the shock passage. Thus, we were able to estimate the time history of local shock-angle $\theta_{BN}$ values and global compression ratios along the shock. 

As an extension to the method presented in \citet{Kozarev:2015}, here we present an analytical DSA model for the possible particle acceleration starting low in the corona, which has been developed to incorporate the remotely observed OCBF properties described above. The unique features of this theoretically simple model are that: 1) It is specifically designed to be run with low coronal observations and model results as input; 2) It provides a fully time-dependent solution for the particle spectra; 3) It has very few free input parameters that are not determined either from direct observations, or from other data-driven model results; 4) It is implemented in the IDL language and is thus easily accessible by the solar and space community. Similar modeling work has been performed previously \citep{Vainio:2008, Battarbee:2013, Afanasiev:2015}. However, these studies used purely analytical expressions for the coronal density and magnetic field. Our intent is to obtain realistic time-dependent spectra of early-stage shock-accelerated Solar Energetic Particles (SEPs) with a physics-based model driven by observations, while retaining the simplicity of the analytical solution which translates into ease of use, applicability to multiple events, and eventually, integration into forecasting tools. We apply the model to explore several idealized cases of proton shock acceleration with typical parameters for the low corona and early-stage eruptions. We also demonstrate the application of the model to a realistic time-dependent scenario.

The structure of the paper is as follows: Section \ref{s2} describes the formulation of the model. Section \ref{s3} discusses the performance of the model and explores the results from a set of idealized run parameters. We present various model results from application to the May 11, 2011 OCBF event in Section \ref{s4}. Section \ref{s5} provides a discussion and summary of the results.

\section{Description of the Model}
 \label{s2}

Consider a large-scale shock front (such as the one discussed in \citet{Kozarev:2015}), which sweeps through the lower corona with speed $V_{shock}(t)$ and strength $r(t)$, crossing at any one time multiple field lines at different angles $\theta_{BN}(t)$. Within a region $\delta x$ surrounding it, the shock will accelerate charged particles of high enough energy, which scatter across it and along each of these field lines with a scattering mean free path, $\lambda_{\parallel}$. The model we developed according to DSA theory \citep{Krymsky:1977, Axford:1977} is based on solving the Parker convection-diffusion equation (Eq. \ref{parkereqn}) for the shock acceleration of ions along individual field lines with varying magnetic field strength, shock speed and strength, and angles $\theta_{BN}$. The amount of field-perpendicular scattering in the model for the parameters used here is negligible - we find ratios of perpendicular to parallel scattering in the range $10^{-12}-10^{-6}$. The solution obtained (Eqns. 8-11) gives, for an initial momentum $p_0$, both the first distribution function ($f_1$) and momentum ($p_1$) values, and and an iterative solution for their values ($f_i$ and $p_i$ at subsequent time steps $i$, separated by the observational cadence $\delta t$ (~12 seconds for SDO/AIA). We obtain the energy-dependent particle injection rate $Q_0(E)$ from a coronal kappa distribution of protons, with $T_{cor}=2\times10^6$~K, $n_{cor}=3\times10^8$~cm$^3$, and $\kappa_{cor}=20$. The solution is found for multiple initial energies, between 10 keV and 1 MeV in this work. The model is run for each individual field line, based on observed and calculated parameters at a single shock-crossing point along it. Flux spectra at each time step are then computed.

We start with injection of particles at rate $Q_0$ and the standard Parker equation \citep{Parker:1965, Jokipii:1966} :
\begin{eqnarray}
\frac{\partial f}{\partial t} + u \frac{\partial f}{\partial x}
     - \frac{\partial }{\partial x}\left( \kappa \frac{\partial f}{\partial x}\right) + \delta x \frac{u_u - u_d}{3}\frac{\partial f}{\partial \ln p}  =  Q_0 \delta x \delta(p-p_0)
\label{parkereqn}
\end{eqnarray}
As usual in diffusive shock acceleration theory \citep{Drury:1983}, we integrate over length $\delta x$, centered on the shock (with $u$ and $d$ as the upstream and downstream indices in the shock frame, respectively):
\begin{eqnarray}
\delta x \frac{\partial f}{\partial t}  + 
\left. \kappa_u \frac{\partial f}{\partial x} \right|_u + 
\frac{\Delta u}{3} \frac{\partial f}{\partial \ln p} = Q_0 \delta(p-p_0),
\end{eqnarray}
where we neglect $\partial f/\partial x$ on the downstream side based on the downstream solution ($f(x >0) = $ constant) of the convection-diffusion equation. We also neglect the $u \partial f/\partial x$ term since $f$ is a continuous solution near the shock and the differential quantity $\delta x$ is vanishingly small. However, we maintain the time-derivative since there are no other quantities giving time differentials. Using the standard upstream solution $f\propto \exp( u_u x/\kappa_u)$, we find
\begin{eqnarray}
\delta x \frac{\partial f_s}{\partial t}  + u_u f_s + \frac{\Delta u}{3} \frac{\partial f_s}{\partial \ln p} = Q_0 \delta(p-p_0).
\label{eq:dif1}
\end{eqnarray}
where $f_s$ is the distribution at the shock. Note that the acceleration rate is given by 
\begin{eqnarray}
\frac{d\ln p}{d\tau} = \frac{\Delta u}{3 \delta x},
\end{eqnarray}
which is consistent  with the theoretically known diffusive acceleration rate \citep{Schwadron:2008}, provided that 
\begin{eqnarray}
\delta x = \frac{\kappa_u}{u_u} + \frac{\kappa_d}{u_d}
\end{eqnarray}
The diffusion coefficients, $\kappa$, can be determined from 
\begin{equation}
	\kappa = \kappa_{\parallel}\cos^2(\theta_{BN}) + \kappa_{\perp}\sin^2(\theta_{BN}),
\end{equation}
where $\kappa_{\parallel}=v\lambda_{\parallel}/3$,   and $\kappa_{\perp}=\kappa_{\parallel}/[1+(\lambda_{\parallel}/r_{g})^2]$.  $v$ is the particle speed, $\lambda_{\parallel}$ is the parallel scattering mean free path in $AU$, $r_{g}$ is the gyroradius of the particle, and $\theta_{BN}$ is the orientation of the upstream magnetic field to the local shock surface normal direction.\\
Taking $y = \ln p$ and $F = f_s \exp(3 u_u y/\Delta u)$, we recast equation (\ref{eq:dif1}) as follows:
\begin{eqnarray}
\delta x \frac{\partial F}{\partial t} +  \frac{\Delta u}{3} \frac{\partial F}{\partial y} = Q_0 \delta( p- p_0).
\end{eqnarray}
This equation can be solved using the method of characteristics (with new variables $t^\prime = t - 3 \delta x y/\Delta u$ and $y^\prime = y$), which yields two possible solutions:
\begin{eqnarray}
f_1 & = & \frac{3 Q_0}{\Delta u p_0} \left(p_1/p_0 \right)^{-\gamma_1} \\
f_i & = & f_{i-1} \left( \frac{p_i}{p_{i-1}}\right)^{-\gamma_i}
\end{eqnarray}
where $\gamma_i = 3 r_i/(r_i-1)$ where $r_i = u_u/u_d $ is the compression ratio at a given step $i$ at the shock crossing point. In this solution 
\begin{eqnarray}
p_1 & = & p_0 \exp\left( \frac{\Delta t \Delta u_1}{ 3 \delta x_1} \right) \\
p_i & = & p_{i-1} \exp\left( \frac{\Delta t \Delta u_i}{ 3 \delta x_i} \right).
\end{eqnarray}

At every time step, at which a field line crosses the shock surface, the change in $f$ is computed by the model, based on the momentum and distribution function value from the previous step. Thus, if a particular line crosses the shock for $N$ total time steps, there will be $(N-1)$ total realizations of the distribution function spectrum according to the scheme above (there must be a minimum of two crossing times to compute a spectrum). The model may be run with multiple initial energies $E_0$, for example taken out of a source distribution specified by the user. Currently, the model does not account for self-generated waves by particles, which are expected enhance the upstream turbulence and increase the acceleration efficiency of quasi-parallel shocks. This will be addressed in future work.

The upstream and downstream diffusion coefficients, $\kappa_u$ and $\kappa_d$, are determined from the shock-to-field angle $\theta_{BN}$, the particle speeds and gyroradii, and the mean free path $\lambda_{\parallel}$. For the current coronal application of the model, we assume that in the low solar corona, the parallel scattering mean free path forprotons is on the order of a convective photospheric granule as a scale-length of the Alfv{\'e}nic fluctuations, i.e. around 1,500~km. Thus, in the runs below, we use $\lambda_{\parallel}=0.022$\rsun. We have made the assumption that photospheric convective motions provide the bulk of wave-like oscillations on the magnetic field lines, which scatter the particles. This argument has been evoked in recent models for the solar wind acceleration and coronal heating \citep{Cranmer:2005}. We will relax this assumption in future work and introduce a more realistic description of the scattering mean free path. We have found that the resulting diffusive scales in the model are larger than a coronal shock approximately 3000-4000~km thick (see Fig. 1 in \citet{Kozarev:2015}), thus satisfying the condition for significant acceleration in DSA.

We assume that the source of seed particles is a coronal proton population, which has the spectral velocity dependence of a kappa distribution \citep{Laming:2013}:
\begin{equation}
f_0 = \frac{n}{2\sqrt{2}(\pi\kappa)^{3/2}v_{th}^{3}}\frac{\Gamma(\kappa)}{\Gamma(\kappa-3/2)}\frac{1}{[1+v^2/2\kappa v_{th}^{2}]^{\kappa}},
\label{eq:kappadist}
\end{equation}
where $v_{th}$ is the thermal speed at temperature $T$, $v$ is the proton speed, $n$ is the coronal density, and $\kappa$ is the defining parameter of the distribution. Thus, we set the injection rate to $Q_0=(f_0\Delta u p_0)/(3)$.

The injection efficiency at the shock for every crossing line and time step is accounted for by applying the injection speed criterion, developed by \citep{Giacalone:2002}. It uses the DSA theory applicability requirement that the diffusive anisotropy of the distribution be a number of order unity \citep{Schwadron:2015}, along with a general expression of the anisotropy, including diffusion parallel and perpendicular to the magnetic field, and drifts. The resulting minimum injection momentum, which charged particles must have in order to be able to catch up with the moving shock, is
\begin{equation}
p_{inj}= m_pu_u\Bigg[1 + \frac{(\kappa_A/\kappa_{\parallel})^2\sin^2\theta_{BN} + \big(1-\kappa_{\perp}/\kappa_{\parallel}\big)^2\sin^2\theta_{BN}\cos^2\theta_{BN}}{\Big((\kappa_{\perp}/\kappa_{\parallel})\sin^2\theta_{BN}+\cos^2\theta_{BN}\Big)^2}\Bigg]^{1/2},
\end{equation}
where $\kappa_A=(pr_g)/(3m_p)$ is the antisymmetric component of the diffusion tensor, containing the effect of drifts. If the input momentum is smaller than $p_{inj}$, the distribution function is not updated at that step.

\section{Verification of the DSA Model: Application to Idealized Coronal Cases}
\label{s3}
To verify the model, and at the same time explore the behavior of DSA-accelerated protons for several typical coronal cases, we devised a grid of input parameters, which were kept constant for the duration of every individual simulation run. Table \ref{testrunstable} shows the parameters chosen for eight different test cases, applicable to low coronal conditions. The runs are denoted with letters A-H. We set shock speeds $V_{shock}=[400, 800]~km/s$, for a slow and fast shock; angle $\theta_{BN}=[5.0, 85.0]~deg$, for a quasi-parallel and quasi-perpendicular shock; and shock ratio $r=[1.3, 2.6]$, for a weak and a relatively strong shock in the corona. The lowest energy of the protons in all cases is $E_0=0.01~MeV$.

\begin{table}[htc]
\centering
\begin{tabular}{c c c c c c c}
\hline
Run Name & $V_{shock}~[km/s]$ & $|B|~[G]$ & $\theta_{BN}~[deg]$ & $r$\\
\hline
A & 800 & 5.0 & 85.0 & 2.6\\
B & 400 & 5.0 & 85.0 & 2.6\\
C & 800 & 5.0 & 5.0 & 2.6\\
D & 400 & 5.0 & 5.0 & 2.6\\
E & 800 & 5.0 & 85.0 & 1.3\\
F & 400 & 5.0 & 85.0 & 1.3\\
G & 800 & 5.0 & 5.0 & 1.3\\
H & 400 & 5.0 & 5.0 & 1.3\\
\hline
\end{tabular}
\caption{The input parameters used for the DSA model's test runs.}
\label{testrunstable}
\end{table}

For each simulation, the DSA model was run for a number of initial energies in the suprathermal range $E=[0.01,1.0]~MeV$, with corresponding distribution function values drawn from the kappa distribution described above. Since the gains in momentum at consecutive steps may be large for the quasi-perpendicular case, as a practical measure we introduce a constant number (100) of substeps between every two consecutive time steps. We use macro time steps of 12~seconds throughout this work, the same as the AIA telescope's cadence. The effective time step for computation is thus 0.12~seconds, while the time resolution of the results presented here remains 12~seconds. This allows to extract time-dependent fluxes with sufficient energy resolution from the model. Each case is run for 40~time steps, or 480~sec. To properly accumulate flux spectra, the resulting time-dependent distribution values are projected onto a regularized energy grid using logarithmic interpolation.


We calculate the initial proton distribution function value corresponding to $0.01~MeV$ protons, assuming a quiet-time coronal distribution at temperature $T=2\times10^6~K$, density $n=3\times10^8~cm^{-3}$, and $\kappa=15$. This value results in a coronal population close to a Maxwellian with very weak suprathermal wings. We chose this value of $\kappa$ such that the distribution value at $0.1~MeV$ would match in magnitude the source population used in previous work \citep{Kozarev:2013}, and determined on the basis of radial scaling of $1~AU$ observations to the solar corona \citep{Dayeh:2009}. We note that strong suprathermal distributions may contribute significantly to the SEP spectra early in events. The proper shape and kappa values must be constrained by comparing model transported spectra at 1~AU to in situ data. This will be pursued in future work. Finally, we have made the simplifying assumption of a constant low coronal solar wind speed of 20~km/s. Solar wind speeds of this order are expected below 2\rsun~\citep{Cranmer:2009}.

\begin{figure}[htc]
\centering
\includegraphics[width=1.0\columnwidth, frame]{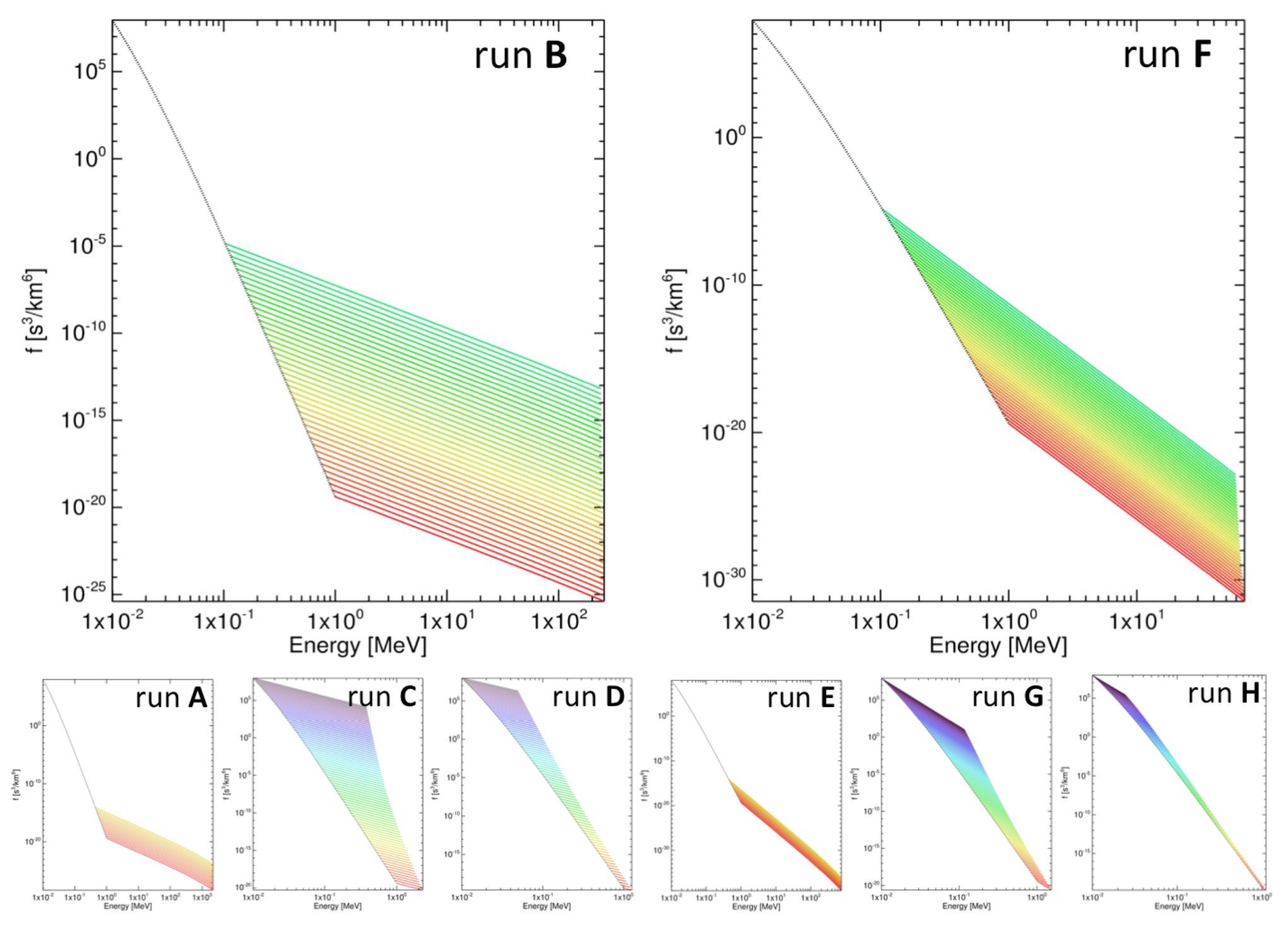}
\setlength{\abovecaptionskip}{2pt}
\setlength{\belowcaptionskip}{-5pt}
\caption{DSA model spectra for a strong (top left panel) and a weak (top right panel) shock case, corresponding to runs B and F from Table \ref{testrunstable}, respectively. Different colors denote different starting energies, with the color coding kept constant among the different panels. The same source kappa distribution with $\kappa=15$ was used in both cases, shown with dotted lines. A grid with 40 initial energies was used in this case. The bottom six panels show the results from the other runs listed in the table. Both vertical and horizontal axis ranges may vary among the panels.}
\label{fig_compare_distributions}
\end{figure}

The resulting distribution function spectra from all eight runs of the DSA model with the parameters listed in Table \ref{testrunstable} are shown in Fig. \ref{fig_compare_distributions}. We show two of the runs, B and F, in larger panels for detail. In the top left panel is the result of run B for a relatively slow and strong quasi-perpendicular shock, which combines spectra from 40 initial energies and distribution values, each spectrum shown in different color. The source kappa distribution is plotted with a dotted line. The individual points on each spectrum are not discernible due to their large number. On the top right panel is the result of shock run F, which has the same speed and orientation as B, but twice as small density compression ratio of 1.3. The other six runs are shown in small panels in the bottom of the figure. 

The comparison between runs B and F reveals the importance of the shock strength for the resulting spectra in DSA. The distribution spectra from run B are much harder than those in run F - as expected in DSA theory - and their lowest values are about 5 orders of magnitude higher. In addition, the shock in run B is able to accelerate protons to almost 200~Mev, compared with 60~MeV in run F. This enhanced acceleration efficiency is also expected. As can be seen, the lowest energies, at which protons experience acceleration in these two runs, are $\sim$100~keV. This is due to the minimum shock injection speed constraint imposed on the solution. The faster and more perpendicular a shock is, the harder it is for the particles to catch up with it and enter the acceleration process. The quasi-parallel shock runs C, D, G, and H are much easier to access by protons below 100~keV, but they are much less efficient at accelerating them to high energies.

\begin{figure}[htc]
\centering
\includegraphics[width=1.0\columnwidth, frame]{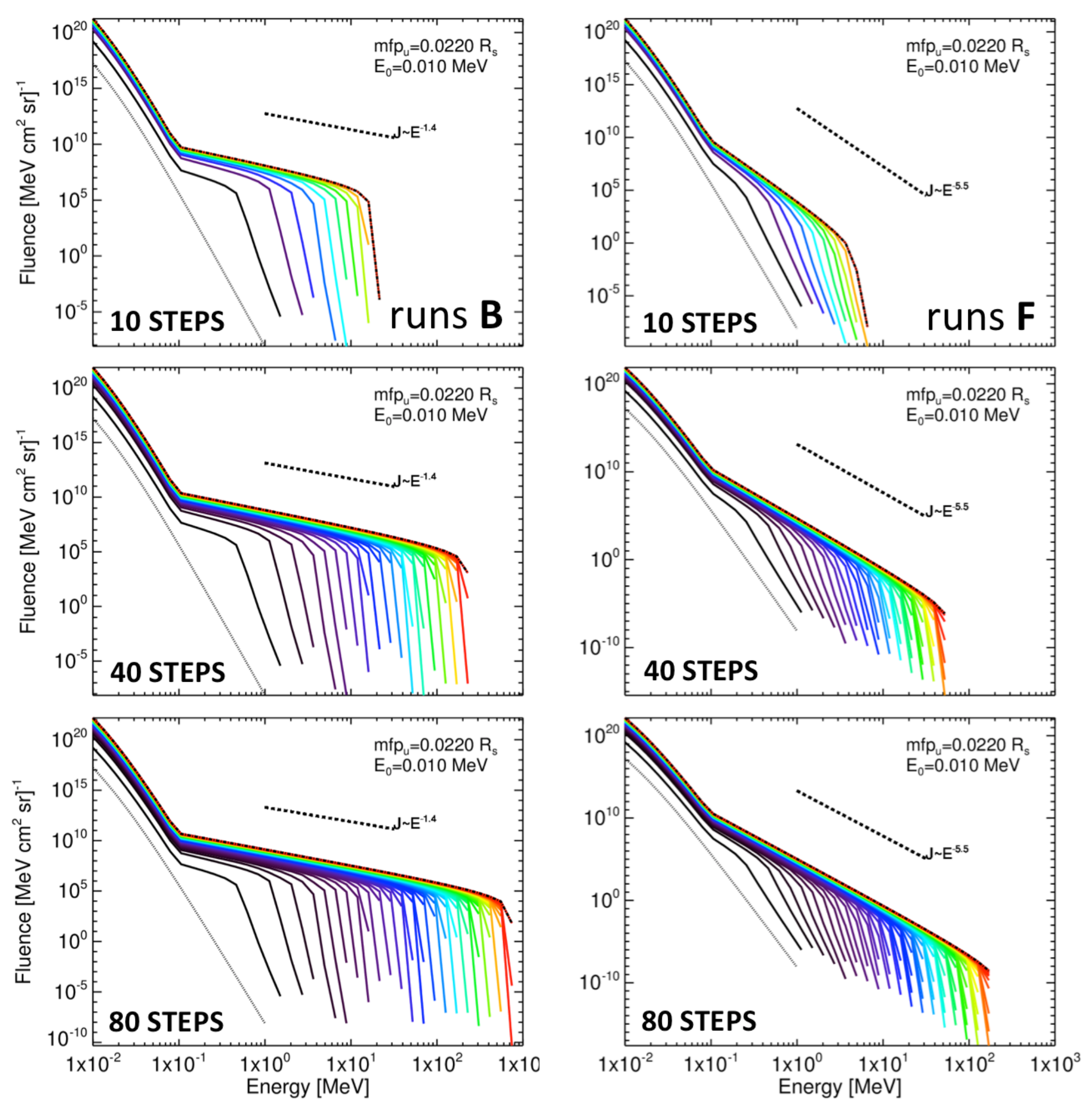}
\setlength{\abovecaptionskip}{2pt}
\setlength{\belowcaptionskip}{-5pt}
\caption{Evolving flux spectra for test runs B and F, and run times of 10 (120~sec), 40 (480~sec), and 80 (960~sec) macro-steps. Different colors denote separate time steps. The dotted lines represent the source input spectrum, and the dashed line shows the final DSA spectra. The convergence to the expected steady state slopes ($\gamma=-1.4$ and $\gamma=-5.5$, respectively; lines shown for comparison) is readily observable, as the spectra evolve over time. The spectra high-energy rollover energy increases with each longer run.}
\label{figcompare_spectra}
\end{figure}

We next look at the flux spectra and how they build up with time. Each panel of Figure \ref{figcompare_spectra} shows the flux spectrum at every time step of the respective DSA model run in a different color, from dark blue to red. The left column is for three runs with run parameters B, the right one for run parameters F. The spectra in the middle panels correspond to the results shown in Fig. \ref{fig_compare_distributions}. We have also added a shorter-duration run of 10 time steps (120~sec) in the top row, and a longer duration of 80 time steps (960~sec) in the bottom row. Again, the input suprathermal spectra, which are the same for all test runs, are shown with dotted lines. The final spectra are overplotted with dashed lines to guide the eye.

The expected steady-state spectrum as a result of diffusive shock acceleration is a power law, with 
\begin{eqnarray}
J & = & J_0 E^{-\alpha},\\
\alpha & = & \frac{1}{2} \frac{r+2}{r-1}.
\end{eqnarray}
In the idealized runs presented here, the power law exponents that correspond to $r=[1.3,2.6]$ are $\alpha=[5.5,1.4]$. For comparison with DSA theory, in Fig. \ref{figcompare_spectra} and other figures below we plot the expected steady-state pre-rollover energy spectral slopes for the prescribed shock strengths as dashed lines. As can be seen, the DSA model performs very well as its results match the theoretically predicted pre-rollover energy spectral slopes.

The comparison of these runs shows well the evolution of the SEP spectra. At the lowest energies, the protons do not have enough energy to undergo the shock acceleration process, so the distribution mimics that of the source. At the higher energies, proton spectra are accelerated significantly in both cases, though to much higher fluxes for the runs B, compared with runs F. As expected in DSA theory, the longer the acceleration proceeds, the more the fluxes are enhanced. Below a characteristic rollover energy the spectra converge to a single power law slope, which is constant for each set of run parameters. The rollover energy increases with consecutive time steps, and the spectra above it soften, in accordance with DSA theory. This behavior is confirmed if we compare the three runs along each column of Fig. \ref{figcompare_spectra} - the spectral shape assumes a steady-state slope, and higher maximum energies, to which the shock accelerates the protons. The maximum energies, to which protons are accelerated are 24.8, 256.0, 775.8~MeV for runs B, and 8.2, 69.6, 228.2~MeV for runs F. To check whether these correspond to the maximum energies expected from DSA, we calculate the acceleration times expected from theory, given by \citep{Jokipii:1987}:
\begin{equation}
\tau_a=\frac{3}{u_u-u_d}\int_{p_0}^{p_f}\Big(\frac{\kappa_u}{u_u} + \frac{\kappa_d}{u_d}\Big)\frac{dp}{p}
\label{accel_time}
\end{equation}
where  $p_0$ and $p_f$ are the starting and final particle momenta, respectively. Upon examination of the model results we concluded that the highest maximum energies are reached when the starting energy is highest -- 1.0~MeV in this work (see Fig. \ref{fig_compare_distributions}). If we set the corresponding momentum as $p_0$, and the final momenta $p_f$ to correspond to the maximum energies from the model runs B and F quoted above, the theoretical acceleration times differ by less than 0.3\% from the total model times for all six cases. Thus, we find that the model agrees very well with theory in this respect.

\begin{figure}[htc]
\centering
\includegraphics[width=1.0\columnwidth, frame]{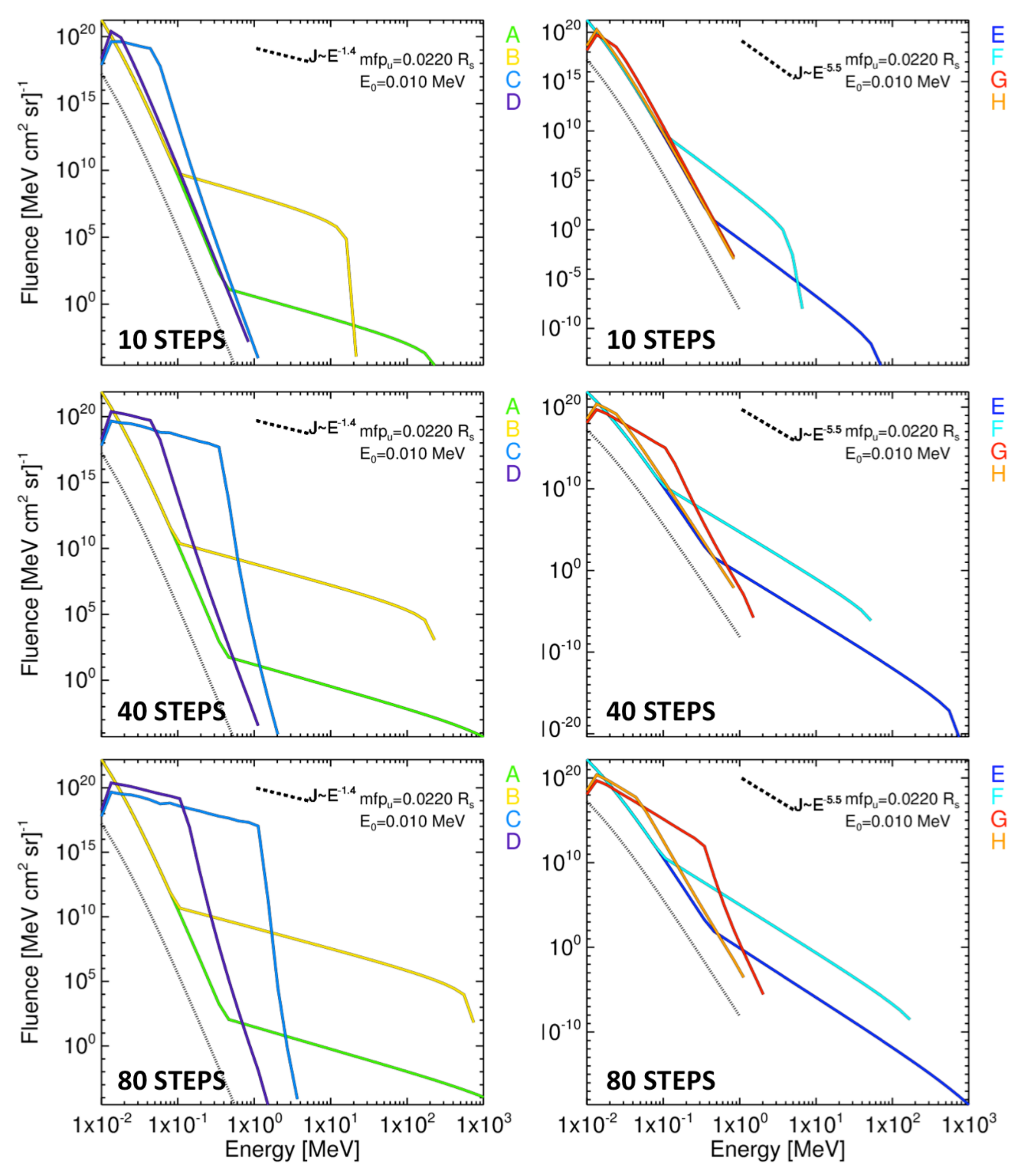}
\setlength{\abovecaptionskip}{2pt}
\setlength{\belowcaptionskip}{-5pt}
\caption{A comparison of the final spectra from verification runs A-H, separated by shock strength (A-D on left and E-H on right), for three run durations (top to bottom). Source spectrum is shown as dotted lines, dashed lines denote steady state. As expected, acceleration is stronger, and rollover energies higher, for higher values of $r$, $\theta_{BN}$, $V_{shock}$. All spectra eventually reach the expected steady state slopes over some portion of the energy range.}
\label{figcompare_final_spectra}
\end{figure}

Next, we compare the final spectra for six sets of runs with all eight parameter combinations listed in Table \ref{testrunstable}, plotted in Figure \ref{figcompare_final_spectra}. The left column shows results for the strong-shock runs A-D, the left column shows results for the weak-shock runs E-H. Every color corresponds to a different parameter combination in Table \ref{testrunstable}. Three different run durations are shown, increasing from top to bottom, analogous to Fig. \ref{figcompare_spectra}. As previously, the source spectra (kept the same throughout) are shown as dotted lines.

The left panels of Fig. \ref{figcompare_final_spectra} show how the final spectra vary for shocks of density jump $r=2.6$. Runs A and B are both quasi-perpendicular, while C and D are both quasi-parallel. In runs A and B the spectra become very hard much earlier than those in runs C and D. The latter two require much longer to start forming the characteristic low-energy steady state slopes. The shock orientation causes the biggest difference in the accelerated proton spectra in the strong shock case, regardless of the large differences in shock speed between runs A-B, and C-D. In fact, for short acceleration durations (top left panel), the spectra from the two quasi-parallel strong shocks C and D are very similar in shape and show only weak enhancement. In a typical low coronal timescale of 8~minutes, both shocks A and B accelerate protons to over 250~MeV, roughly two orders of magnitude higher energies than the efficiency of shocks C and D. However, the slower shock B has produced consistently higher fluences due to the restriction to lower energy protons to enter the faster shock A.

In the right panels of Fig. \ref{figcompare_final_spectra}, the results of the runs for the relatively weak shocks ($r=2.6$) show very similar trends. The spectra are differentiated mainly by the shock orientation, with runs E and F producing the most proton acceleration. The protons accelerated in faster shocks are also the first to reach higher energies, though again we note that the slower quasi-perpendicular shock F has produced higher fluences than those of the faster shock E.

Overall, the comparison shows that strong shocks produce larger fluxes than weak shocks at energies above 1~MeV. Due to the much longer acceleration timescales for quasi-parallel shocks, the final fluences of runs C, D, G, and H are quite similar for the 10- and 40-time step runs, and only differ significantly for the longest duration set of runs. An interesting find is that beyond 1~Mev, the slow quasi-perpendicular shocks (B, F) produce larger overall fluences than the fast ones (A, E). The trend is reversed for the quasi-parallel shocks - the rollover energies achieved by the slow ones (D, H) is low enough that the fast ones (C, G) overtake them in terms of fluences at all energies above 1~MeV. This may have implications for future in situ observations by Solar Probe Plus.

Table \ref{comparetestruns} summarizes the results from the eight sets of runs, listing for each the fitted pre-rollover spectral slope $\alpha$, the rollover energy, the maximum energy to which protons are accelerated, and the fluence at that energy. The rollover energy was determined manually based on the deviation of the power law fit from the model spectrum at the high-energy end. We have chosen a manual method, since some of the generated spectra do not lend well to automatic fitting due to apparent lack of a power-law portion. A dash replaces the spectral slope of spectra with no steady-state part. The slopes of the spectra with more acceleration are slightly softer than the expected steady state slope, which reflects a more gradual departure from the steady-state power law shape than in the cases with less acceleration.


\begin{table}[htc]
\setlength{\tabcolsep}{14pt}
\footnotesize{
\centering
\begin{tabular}{c c c c c c c}
\hline
~& $A_{10}$ & $A_{40}$ & $A_{80}$ & $B_{10}$ & $B_{40}$ & $B_{80}$\\
\hline
$\alpha$ & 1.65 & 1.69 & 1.72 & 1.70 & 1.61 & 1.61\\
$E_{roll}$ & 39.57 & 300.40 & 353.28 & 3.93 & 30.61 & 85.49\\
$E_{max}$ &279.52 & 2179.65 & 5030.16 & 24.81 & 256.07 & 775.82\\
$J_{max}$ &2.70e-05&5.08e-05&1.09e-04&1.22e-04&1.16e03&6.55e01\\
\hline
\hline
~& $C_{10}$ & $C_{40}$ & $C_{80}$ & $D_{10}$ & $D_{40}$ & $D_{80}$\\
 \hline
$\alpha$ & 1.40 & 1.34 & 1.36 & - & 1.39 & 1.40\\
$E_{roll}$ & 0.04 & 0.34 & 1.12 & 0.02 & 0.04 & 0.11\\
$E_{max}$ & 1.28 & 2.30 & 4.14 & 1.06 & 1.26 & 1.55\\
$J_{max}$ & 9.09e-05 & 8.01e-05 & 7.61e-05 & 1.51e-03 & 3.62e-04 & 2.82e-05\\

\hline
\hline
~& $E_{10}$ & $E_{40}$ & $E_{80}$ & $F_{10}$ & $F_{40}$ & $F_{80}$\\
\hline
$\alpha$ & 5.80 & 5.81 & 5.90 & 6.00 & 5.74 & 5.69\\
$E_{roll}$ & 25.34 & 197.59 & 304.49 & 2.62 & 20.41 & 57.74\\
$E_{max}$ & 76.11 & 757.11 & 1995.87 & 8.22 & 69.65 & 228.18\\
$J_{max}$ & 1.88e-14 & 4.19e-21 & 2.63e-19 & 1.02e-08 & 7.51e-07 & 2.96e-09\\
\hline
\hline
~& $G_{10}$ & $G_{40}$ & $G_{80}$ & $H_{10}$ & $H_{40}$ & $H_{80}$\\
\hline
$\alpha$ & - & 5.50 & 5.48 & - & 4.82 & 5.40\\
$E_{roll}$ & 0.03 & 0.11 & 0.35 & 0.01 & 0.02 & 0.04\\
$E_{max}$ & 1.12 & 1.54 & 2.20 & 1.03 & 1.12 & 1.24\\
$J_{max}$ & 1.78e-03 & 1.71e-06 & 2.81e-06 & 1.11e-03 & 7.51e-03 & 2.88e-04\\
\hline
\end{tabular}
\caption{A comparison between the output parameters of the time-invariable test runs of the DSA model.}
\label{comparetestruns}
}
\end{table}

The results presented above not only serve to verify the performance of the DSA model, but also illustrate the proton acceleration to be expected by shocks of different strengths and orientations, for typical low coronal conditions and propagation timescales. The model results show that strong quasi-perpendicular shocks in the low corona can accelerate proton seed populations to energies well above 100~MeV, provided that the favorable conditions are sustained over the transit times on the order of 10~minutes. However, such extreme prompt flux enhancements connected to coronal shocks are rarely observed. Most shocks steepen gradually and increase in strength according to local conditions, but should still produce some SEPs. In the next section, we show that in real events the relevant acceleration parameters at the shock crossings may vary continuously as the disturbance propagates, requiring the detailed time-dependent, data-driven approach that our DSA model provides.

\section{Application to the May 11, 2011 OCBF event}
\label{s4}
We have applied the DSA model to the recently studied OCBF event of May 11, 2011 (See \citet{Kozarev:2015} for details on the event). We will refer to it as the `May11' event. We used the kinematics calculations and modeled density changes from that study as time-dependent inputs to the DSA model. The upstream and downstream speeds were obtained from the AIA observations of shock surface kinematics discussed in Sect. \ref{s1}. The time-dependent magnetic field magnitude, density compression ratio, and shock angle were determined from observed kinematics, PFSS and DEM model results. The high-resolution PFSS model, generated for the May 11 event, yielded 176 magnetic field lines that intersected the shock surface for three or more time steps. The DSA model was run on all of them. The region of interaction between the shock and magnetic field lines is approximated as a point, which allows us to extract single values of the relevant parameters for each time of interaction. We should point out that since our model is run on multiple lines, the acceleration times for different magnetic field lines will vary due to the different parameters at the line crossings. In addition, since the model is time-dependent, the theoretical acceleration time (Eq. \ref{accel_time}) could only give an approximation to the actual values in the model.

\begin{figure}[htc]
\centering
\includegraphics[width=1.0\columnwidth, frame]{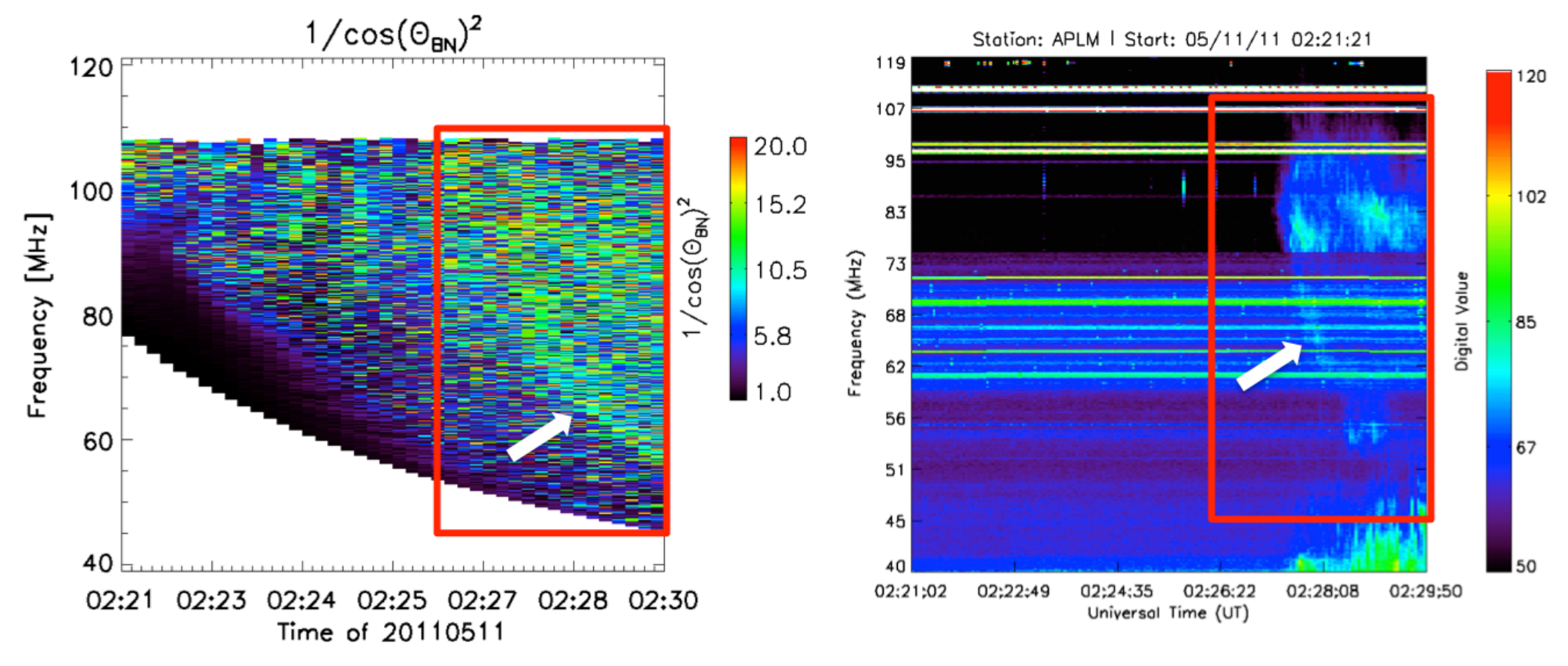}
\setlength{\abovecaptionskip}{2pt}
\setlength{\belowcaptionskip}{-5pt}
\caption{A comparison between the modeled shock acceleration efficiency in time and frequency, and the observed type II radio burst by Learmonth solar observatory. Left panel: a reconstructed spectrogram showing the maximum values of the parameter $1/\cos^2(\theta_{BN})$ - proportional to the rate of particle acceleration - as a function of time and frequency, assuming a scaled \citet{Mancuso:2008} radial density model. Right panel: Observed radio emission during the beginning of the May11 event shows the onset of a type II burst. The red box has the same extent in both time and frequency in the two panels, and shows the period of high values of $1/\cos^2(\theta_{BN})$, which coincide with the onset of the type II burst. The white arrows point to the distinct slowly drifting bands seen in both panels.}
\label{figradioshockcompare}
\end{figure}

Using the times and radial heights of the shock-crossing points, we have reconstructed a spectrogram of the shock acceleration efficiency. Figure \ref{figradioshockcompare} shows a comparison between (on the left) the parameter $1/\cos^2(\theta_{BN})\sim d\ln p/d\tau$, which is proportional to the acceleration rate in DSA  (see Eqs. 4-6), and (on the right) the observed type II radio emission intensity by the Learmonth solar radio observatory (Learmonth, Australia), during the period in which AIA observed the event. We calculated the frequency dependence on the basis of a 0.5-fold coronal radial density model \citep{Mancuso:2008}, with the scaling factor chosen to match the radio observation frequencies during the period of increased emission. The time resolution is 12~seconds, and the frequency resolution - 0.1~MHz. The color coding in the left panel corresponds to the maximum value of $1/\cos^2(\theta_{BN})$ in each of the time-frequency bins. The red box denotes a period of enhanced value of this parameter towards the end of the event as observed in the AIA FOV. A box of the same extent in time and frequency is drawn on the right panel with actual observations. The horizontal lines on that spectrogram are terrestrial radio sources. It can be readily observed that the beginning of the observed type II event coincided with the period of high values of $1/\cos^2(\theta_{BN})$. There is a particularly good temporal and frequency agreement for the easily discernible, slowly drifting band between 90 and 55~MHz (marked with white arrows on the two panels). This comparison to the observed type II radio burst gives us confidence in claiming that a shock was observed in the AIA FOV, which was already accelerating particles. In addition, it points to the usability of the PFSS model for modeling the magnetic field in the low corona.

\begin{figure}[htc]
\centering
\includegraphics[width=1.0\columnwidth, frame]{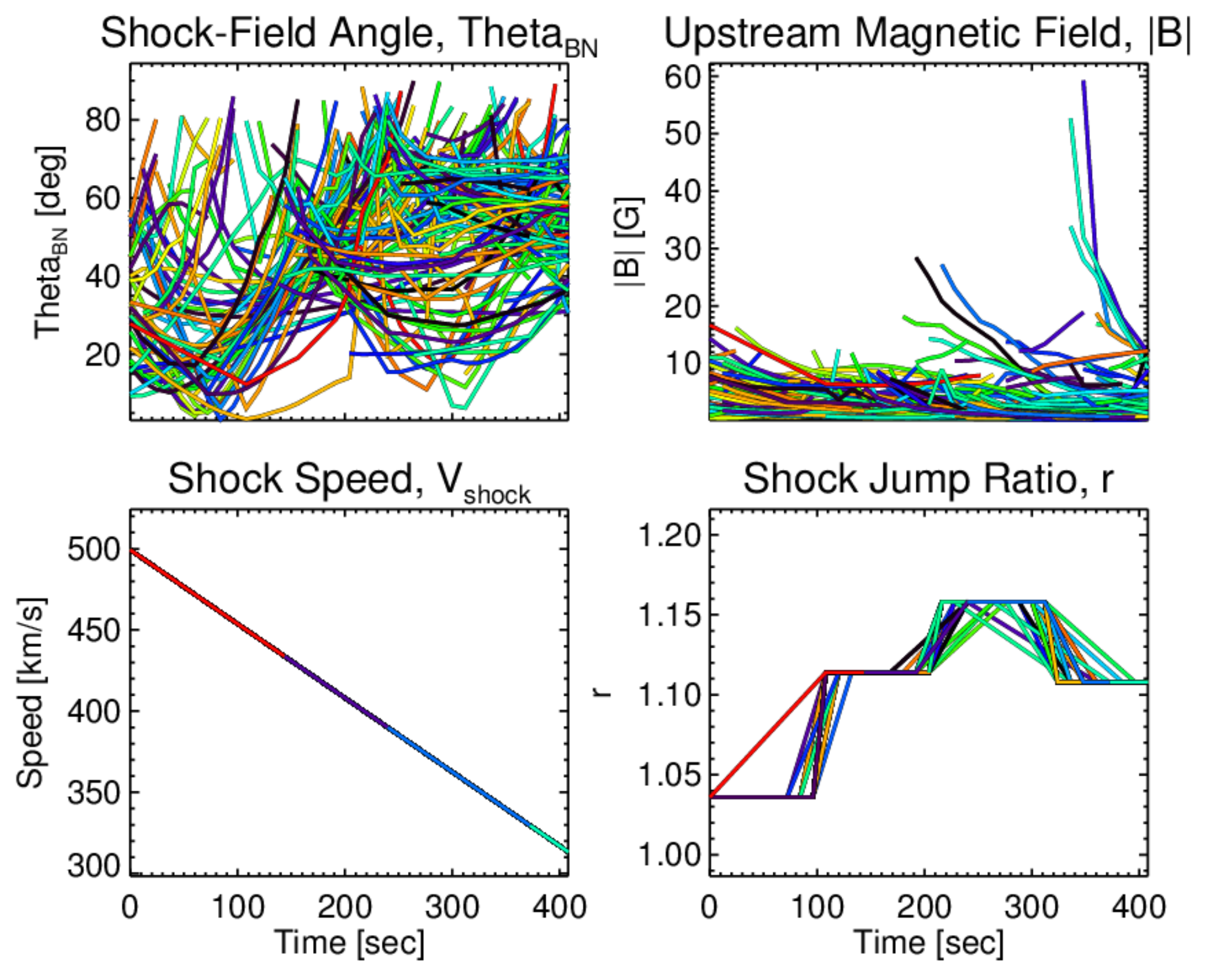}
\setlength{\abovecaptionskip}{2pt}
\setlength{\belowcaptionskip}{-5pt}
\caption{Time series of parameters used in the DSA model, calculated for the shock-crossing points of 176 field lines during the May 11 OCBF coronal passage. Clockwise, from top left: $\theta_{BN}$, $|B|$, $r$, $V_{shock}$. The time history of the crossings on each point is shown in a different color. On the X-axis is time since the beginning of the event.}
\label{figshockinfo}
\end{figure}

Figure \ref{figshockinfo} shows the evolution of the four main relevant parameters of the shock observed during the May 11 event immediately upstream of the shock-crossing locations of all 176 field lines used to model the proton acceleration. The panels show, clockwise from top left, angle $\theta_{BN}$, magnetic field magnitude $|B|$, density compression ratio $r$, and shock speed $V_{shock}$. Time in seconds is on the x-axis. The time history of every shock crossing location is in different color. It is immediately obvious that the $\theta_{BN}$ angle values exhibit significant variation over the period of shock crossing due to the particular crossing geometries between the spherical shock surface and the loop-like field lines. The values of the $\theta_{BN}$ angle vary between 3.3\degrees~and 89.9\degrees.This variation changes the shock acceleration efficiency significantly, as was shown in Sec.~\ref{s3}. The magnetic field magnitude at the shock-crossing points also shows significant variation, and in some cases reaches almost 60~G. Overall, the values vary between 0.2~G and 59.2~G.

The density compression ratio, $r$, is perhaps the most important parameter, and also the most difficult to determine from observations. In the absence of sophisticated MHD simulations combined with comparisons of the forward-modeled EUV emissions to observations \citep{Downs:2012}, a popular alternative to obtaining observations-based density changes in coronal shocks is the DEM modeling approach \citep{Kozarev:2011, Vanninathan:2015}. In this method, the changes in density are determined by comparing line-of-sight-integrated DEMs from EUV observations before and during the event. The shock jump density ratios used here were determined by \citet{Kozarev:2015} by averaging pixel values in several regions along the radial direction of propagation of the OCBF. Thus, they do not correspond to individual crossing points; we will improve the density determination in future work. The values of $r$ are quite low for this event, representing a consistently weak increase in density of only 6\%-16\% along the nose of the shock. This may be a systematic issue due to the specifics of the DEM method: line-of-sight averaging of the emission, and the assumption that most plasma along the line of sight is at a limited temperature range. Future work using a variety of DEM models, and comparisons to radio observations will help improve this method. In any case, given the range of density compression ratios, we do not expect significant acceleration for this event. Nevertheless, the density compression ratio exhibits a slight increase over time, which may signify overall strengthening of the shock. 

Finally, the shock speed time history is shown in the bottom left panel of Fig.~\ref{figshockinfo}, varying linearly between 313~km/s and 499~km/s. The speed profile is based on a second-order polynomial fit to the radial positions of the shock front measured along the direction of movement of the shock nose, so the time series of all crossings overlap. In future work, we will improve the measurements for individual crossing points.

\begin{figure}[htc]
\centering
\includegraphics[width=1.0\columnwidth, frame]{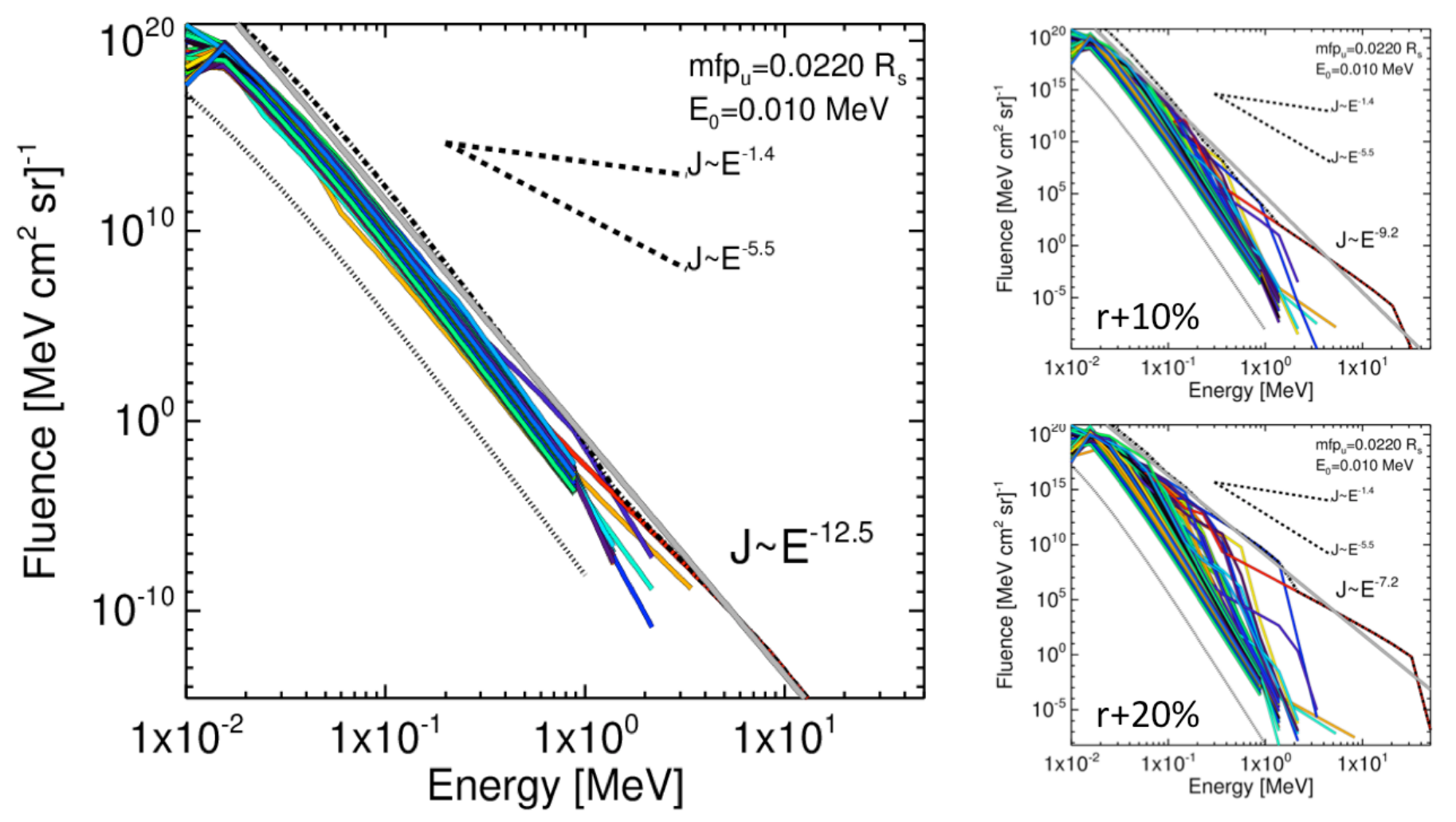}
\setlength{\abovecaptionskip}{2pt}
\setlength{\belowcaptionskip}{-5pt}
\caption{Proton fluence spectra for the May 11, 2011 shock event. Left panel: fluence spectra for each of the 176 shock-crossing lines are plotted with a different color. The source spectrum, extending to 1 MeV, is shown with a dotted line. A power law fit to the sum of all the spectra (dot-dashed line) is plotted with a grey continuous line, and the slope value of -12.5 is shown. The slopes corresponding to the steady-state modeling cases are also shown for comparison. The right panels show the same plot but when the model is run with a compression ratio augmented by 10$\%$ (top) and 20$\%$ (bottom). The overall spectrum slopes rise to -9.2 and -7.2, and the highest energies reached increase to 30 and 50~MeV, respectively.}
\label{fig_fluences_may11}
\end{figure}

For the modeling of proton acceleration, we keep most model parameters as in the steady-state cases in \ref{s3}, and only vary the four main parameters, $\theta_{BN}$, $|B|$, $r$, $V_{shock}$ according to the observations/data-driven model results (Fig. \ref{figshockinfo}). Figure \ref{fig_fluences_may11}, left panel, shows the resulting final proton spectra for all 176 lines modeled with the DSA code. For this event with weak compression early on, most of the spectra do not exhibit enhancement or extend above 1~MeV, with several exceptions. This is mostly due to the very low values of $r$. Several spectra extend to 2-3~MeV, and one spectrum shows accelerated particles to above 10~MeV. We also fit power laws to the total spectrum, obtained by summing individual line fluences. We use these power law fits only as a reference for the overall level of acceleration, as the model spectra are not steady-state, they have high-energy rollovers at different energies, and may be dominated by only a few lines at the highest energies. For the May11 event, the grey solid line in Fig. \ref{fig_fluences_may11} shows the power law fit and the resulting slope of -12.5. This low exponent is indicative of very little acceleration.

We note that the density compression ratios used in the model may be an underestimation, as the DEM method has previously shown lower results than a method using radio observations. A study of the June 13, 2010 OCBF event by \citet{Ma:2011}, using observations of metric type-II radio bursts with band splitting, found the density increase to be 36\% higher than that obtained with DEM analysis by \citet{Kozarev:2011}. On the other hand, such weak density compressions are common for CBF events \citep{Vanninathan:2015}. Thus, we have re-run the model for two cases with slightly increased density compression ratios. The right top and bottom panels of Fig. \ref{fig_fluences_may11} show the resulting spectra if the density compression ratios are increased by 10\% ($1.16<r<1.28$) and 20\% ($1.27<r<1.39$), respectively. This leads to a rise of the slopes of the power law fits to the total spectrum to -9.2 and -7.2, and the highest energies reached to 30 and 50~MeV, respectively. The uncertainty associated with estimating this sensitive parameter must be addressed by improving the DEM method in the future, and combining it with estimations from radio observations. Furthermore, while the PFSS model is acceptable for the low corona, its applicability is limited to events westward of the central meridian (such as this one), and it may be better to use synoptic global MHD model results for detailed future studies. 

\begin{figure}[htc]
\centering
\includegraphics[width=1.0\columnwidth, frame]{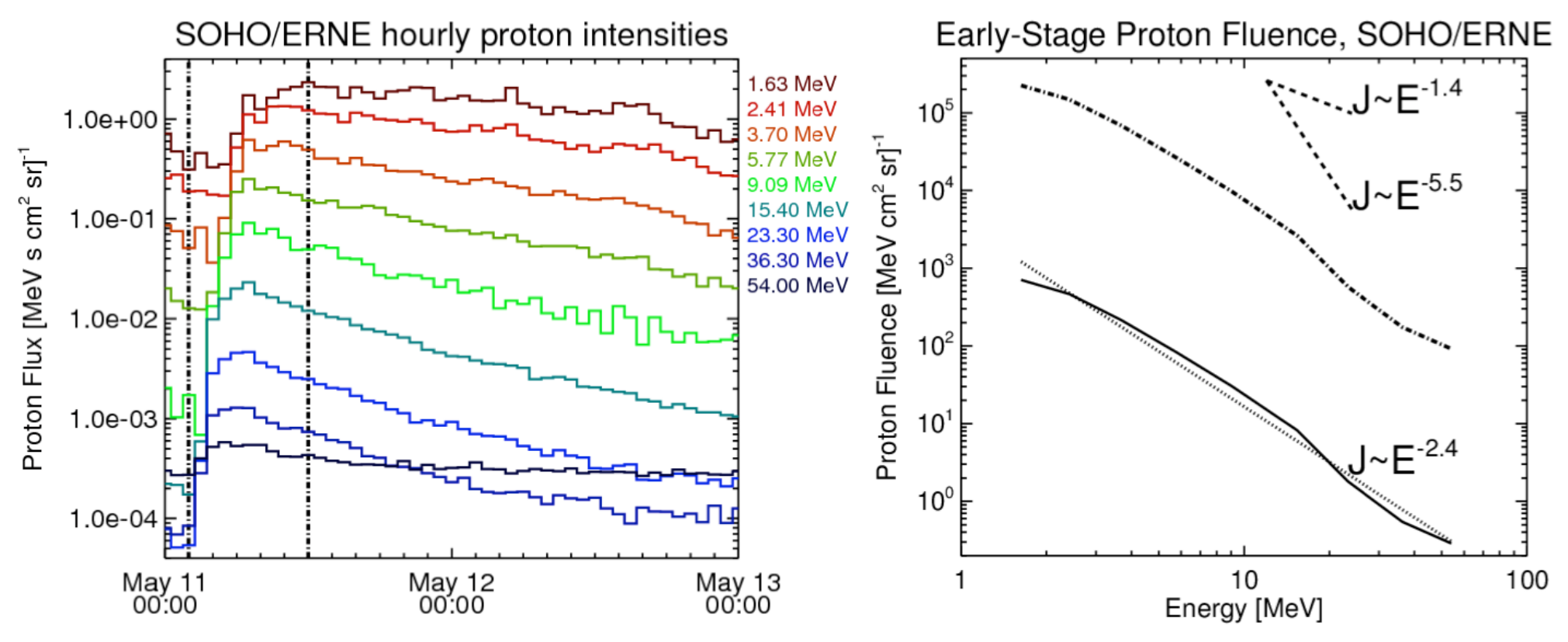}
\setlength{\abovecaptionskip}{2pt}
\setlength{\belowcaptionskip}{-5pt}
\caption{Observed SEP fluxes at SOHO/ERNE during the May11 event. Left panel: hourly proton fluxes in the nine ERNE channels with flux enhancement, observed during May 11 and 12. The two vertical dot-dashed lines denote the start of the OCBF event on the Sun and the end of the SEP event onset, respectively. Right panel: The integrated fluence during the SEP event onset stage is shown with a solid line, along with a power-law fit (dotted line). The slope of the fit is -2.4. The dot-dashed line denotes the fluence rescaled for a 10-minute period and back to the Sun (see text).}
\label{fig_erne_protons}
\end{figure}

The May11 event produced SEPs at 1 AU up to $\sim50$~MeV. Figure \ref{fig_erne_protons}, left panel, shows the hourly-averaged proton fluxes observed by the SOHO/ERNE \citep{Torsti:1995} instrument for May 11 and 12. Prior to the event the fluxes at the lower energies below 10~MeV were already elevated, while for the higher energies this was a larger increase. No increase was observed for energies above 55~MeV. The velocity dispersion is visible, indicating the fluxes were injected near the Sun. The first dot-dashed vertical line denotes the start of the OCBF event in the solar corona, while the second one is positioned at what we define as the end of the event onset - the time when the lowest-energy fluxes peaked. The solid line in the right panel shows the integrated proton fluence spectrum for the onset of the event - the period between the two dot-dashed lines. The period was chosen to capture the high energy fluences at the onset of the event. This approach has the shortcoming that lower energies are underrepresented in the fluences, and thus the calculated spectrum is harder than if we had chosen a period more representative of the steady state. To properly address this issue, in future work we will model the interplanetary transport for direct comparison.

We fit a power law to this onset spectrum, and find a slope of -2.4 -- much harder than the modeled one. This relatively hard slope compared with the model results can be explained with two effects -- 1) the bulk of the acceleration occurred beyond the domain of our model (beyond 1.5\rsun); and 2) the observed fluence spectrum presented here combines varying acceleration efficiency with transport effects between the Sun and 1~AU. Since the higher energy particles arrive first at the spacecraft, and are in general scattered less than the lower energy particles, the observed spectrum of the event onset will be harder than for the entire event. Thus, the transport from the acceleration region to the observer will not necessarily maintain the same power law as seen at the shock. For a proper and detailed comparison to data, in future work we will model the shock acceleration and heliospheric proton transport using a numerical model. For a qualitative comparison of the fluence values here, we have scaled the spectrum back to 1.5~\rsun, assuming a radial distance power law scaling with a slope -1.93 \citep{Dayeh:2010}. We have also rescaled the fluences to a 10-minute period. The result is the dot-dashed line on the right panel. Based on this rescaling, the fluence near 2~MeV is about 10 orders of magnitude larger than the fluence in the left panel of Fig. \ref{fig_fluences_may11}, but is less than two orders of magnitude from the 2~MeV fluence in the right lower panel in that figure. The same is true for the fluences at 10~MeV. We defer a detailed quantitative comparison to in situ observations for a future study.

\begin{figure}[htc]
\centering
\includegraphics[width=0.75\columnwidth, frame]{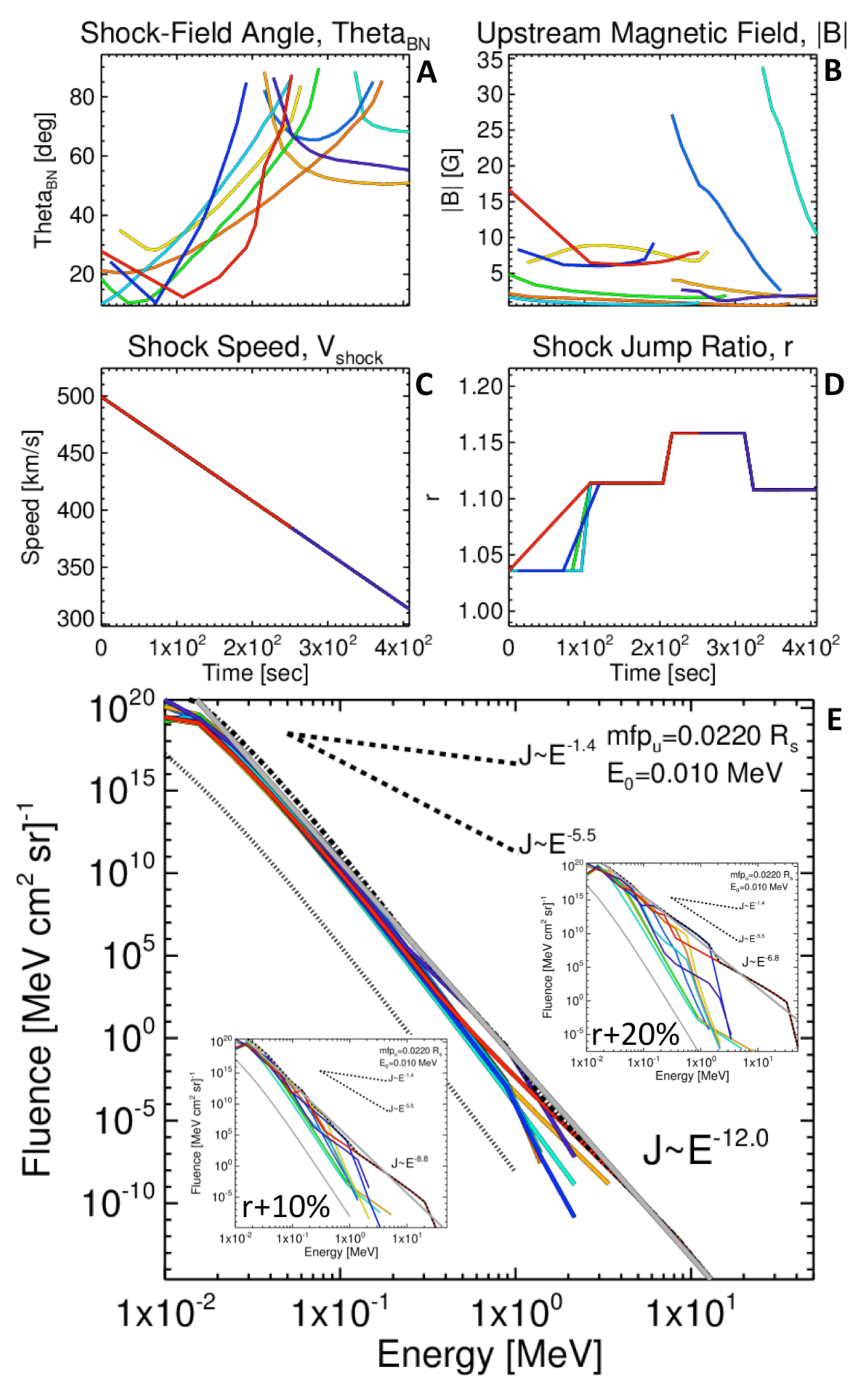}
\setlength{\abovecaptionskip}{2pt}
\setlength{\belowcaptionskip}{-5pt}
\caption{(Panels A-D) Time-dependent input parameters for the nine lines with strongest shock acceleration and one with weak acceleration - similar to Fig. \ref{figshockinfo} and Fig. \ref{fig_fluences_may11}. (Panel E) Final fluence spectra for these 10 lines, with insets showing spectra if shock strength is increased by 10\% (lower left) and 20\% (upper right), with total spectral slopes of -8.8 and -6.8, respectively. The total fluence spectrum (dot-dashed line) and a power-law fit (full grey line) to it is shown in each panel.}
\label{figshockinfosub}
\end{figure}

Finally, we take a closer look at the parameters and results for the lines, along which most significant acceleration occurred. Figure \ref{figshockinfosub} shows input parameters (panels A-D) and model output (panel E) for nine field lines with relatively significant acceleration, and one line with very little proton acceleration. The lines were chosen on the basis of both maximum energy reached and fluence enhancement at spectral energies above 2~MeV. Based on the input parameters, more efficient acceleration can be traced to the combination of higher $\theta_{BN}$ and $r$ values later in the event. The most acceleration occurred at the red-colored line, for which the shock-crossing point experienced gradual increases in $\theta_{BN}$ up to $\sim$85\degrees, while the magnetic field magnitude was relatively high early in the event and later decreased. 
This underscores the importance of having temporal information on all relevant parameters for particle acceleration, not just $\theta_{BN}$. The insets in panel E of Fig. \ref{figshockinfosub} show the resulting fluence spectra if the density compression ratios are increased by 10\% (lower left) and 20\% (upper right). The fitted power law spectral slopes are -12.0 for the nominal run, -8.8 for the r+10\% run, and -6.8 for the r+20\% run. These values are higher than the corresponding values when all lines are included (Fig. \ref{figshockinfo}) - we think this is due to the low-energy contributions from the lines with little acceleration. The fitted slopes of the source spectrum (dotted lines in the plots) are -12.9 for the entire energy range (0.01-1.0~MeV), and -13.7 for the range 0.1-1.0~MeV. These are lower than the model slopes, signifying some particle acceleration occurred in all model runs.



\section{Summary}
\label{s5}
We have presented a new analytic model based on DSA theory, specifically developed to take as input remote solar observations and data-driven model output, which solves for the coronal charged particle acceleration by large-scale CME-driven shocks. Our model uses time-dependent estimates of shock speed $V_{shock}$, density jump ratio $r$, magnetic field strength $|B|$ and shock angle $\theta_{BN}$, for multiple shock-crossing field lines. The model includes a consideration of the minimum shock injection speeds, and a source population drawn out of a coronal $\kappa$-distribution.

We performed a model verification by running it for a grid of steady-state typical coronal shock parameter values. We have shown the model to reproduce theoretical DSA results for conditions approaching a steady state. As expected, acceleration is stronger, and rollover energies higher, for higher values of $r$, $\theta_{BN}$, and $V_{shock}$. All spectra eventually reach the theoretical steady-state slopes over some portion of the energy range. We find that even shocks with compression ratio as low as 1.3 can accelerate protons up to 100~MeV in low coronal conditions if they are sufficiently fast and quasi-perpendicular. 

We applied the model to the OCBF event of May 11, 2011. That event showed a dome-like compression front expanding at higher-than-average speeds, which however showed only mild density enhancements. We have suggested in a previous study \citep{Kozarev:2015} that the observed front was a compressive wave, which steepened to a true discontinuity only as it was about to exit the AIA FOV. Nevertheless, we found that in the low coronal stage of the May11 event, protons may have been accelerated up to at least 10~MeV, and possibly higher energies, if the compression ratios obtained from DEM modeling are considered as lower limits. This shows that significant SEP acceleration, albeit weak, likely occurred for the May11 event below 1.5\rsun. A corresponding, relatively small SEP event was observed at SOHO, which had the profile of a typical western event, and proton fluxes enhanced up to $\sim$50~MeV. A qualitative comparison with the model proton spectra near the Sun leads us to conclude that since the simulated spectra are quite soft and limited in energy extent (and the event was accompanied by a very weak B8.1 flare), the bulk of proton acceleration must have occurred above 1.5\rsun~at the evolving CME shock. We suggest that the initial low coronal acceleration may also have formed the source population for further acceleration higher in the corona.

With this study we have taken a first step in using direct observations of shocks and compressions in the innermost corona to predict the onsets and intensities of SEP events. The model has high potential for space weather prediction and scientific support for interpretation of energetic particle observations by the upcoming Solar Probe Plus and Solar Orbiter missions. In future work, we will improve the density change estimation by calculating the compression at every shock-crossing point, and the magnetic field estimation by including synoptic 3D MHD simulation results. In addition, we will extend the model domain higher in the corona by measuring shock kinematics and compression using white-light coronagraph observations. This will allow us to study the locations of bulk acceleration, and the relevant coronal conditions. The DSA model presented here will be used for studying the early stages of coronal shock particle acceleration of a sample of over 60 OCBF events from a catalog being compiled at SAO under the Coronal Analysis of Shocks and Waves (CASHeW) framework.


\acknowledgements
KAK acknowledges support through a NASA GI grant. We would like to thank Maher Al Dayeh for providing the SOHO/ERNE data. 

\bibliography{corwav_accel}
\end{document}